\documentclass[twocolumn,showpacs,preprintnumbers,amsmath,amssymb]{revtex4}

\usepackage{graphicx}% Include figure files
\usepackage{dcolumn}% Align table columns on decimal point
\usepackage{bm}% bold math
\usepackage{multirow}
\usepackage{mathtools}
\usepackage{amsmath}
\usepackage{array}
\usepackage{siunitx} % Required for alignment
\usepackage{booktabs}
\usepackage{xcolor}
\usepackage{subfig}
\usepackage{float}
\newcolumntype{P}[1]{>{\centering\arraybackslash}p{#1}}
\newcommand{\ra}[1]{\renewcommand{\arraystretch}{#1}}

\AtBeginDocument{
\heavyrulewidth=.08em
\lightrulewidth=.05em
\cmidrulewidth=.03em
\belowrulesep=.65ex
\belowbottomsep=0pt
\aboverulesep=.4ex
\abovetopsep=0pt
\cmidrulesep=\doublerulesep
\cmidrulekern=.5em
\defaultaddspace=.5em
}

\begin{document}

\preprint{}

\title{Three-state Opinion Dynamics for Financial Markets on Complex Networks}

\author{Bernardo J. Zubillaga}
\affiliation{Center for Theoretical Biological Physics and Department of Physics, Northeastern University, Boston, MA 02115, USA.}
\affiliation{Center for Polymer Studies and Department of Physics, Boston University, Boston, MA 02215, USA}

\author{Mateus F. B. Granha}
\affiliation{F\'isica de Materiais, Universidade de Pernambuco, Recife, PE 50720-001, Brazil}
\affiliation{Departamento de F\'{\i}sica, Universidade Federal de Pernambuco, Recife, PE 50670-901, Brazil}

\author{Andr\'e L. M. Vilela}
\affiliation{F\'isica de Materiais, Universidade de Pernambuco, Recife, PE 50720-001, Brazil}
\affiliation{Departamento de F\'{\i}sica, Universidade Federal de Pernambuco, Recife, PE 50670-901, Brazil}
\affiliation{Data Science and Analytics, SUNY Polytechnic Institute, Utica, NY 13502, USA}
\affiliation{Center for Polymer Studies and Department of Physics, Boston University, Boston, MA 02215, USA}

\author{Chao Wang}
\email{chaowanghn@vip.163.com}
\affiliation{College of Economics and Management, Beijing University of Technology, Beijing, 100124, China}

\author{Kenric P. Nelson}
\affiliation{Photrek LLC, Watertown, MA 02472, USA}

\author{H. Eugene Stanley}%
%\email{hes@bu.edu}
\affiliation{Center for Polymer Studies and Department of Physics, Boston University, Boston, MA 02215, USA} 

\date{\today}

\begin{abstract}

This work investigates the effects of complex networks on the collective behavior of a three-state opinion formation model in economic systems. Our model considers two distinct types of investors in financial markets: noise traders and fundamentalists. Financial states evolve via probabilistic dynamics that include economic strategies with local and global influences. The local majoritarian opinion drives noise traders' market behavior, while the market index influences the financial decisions of fundamentalist agents. We introduce a level of market anxiety $q$ present in the decision-making process that influences financial action. In our investigation, nodes of a complex network represent market agents, whereas the links represent their financial interactions. We investigate the stochastic dynamics of the model on three distinct network topologies, including scale-free networks, small-world networks and Erd{\"o}s-R\'enyi random graphs. Our model mirrors various traits observed in real-world financial return series, such as heavy-tailed return distributions, volatility clustering, and short-term memory correlation of returns. The histograms of returns are fitted by coupled Gaussian distributions, quantitatively revealing transitions from a leptokurtic to a mesokurtic regime under specific economic heterogeneity. We show that the market dynamics depend mainly on the average agent connectivity, anxiety level, and market composition rather than on specific features of network topology.

\end{abstract}

%\pacs{87.23.Ge Dynamics of social systems, 05.10.Ln Monte Carlo methods, 64.60.Cn Order-disorder transformations}


\maketitle

\section{Introduction}

Network science has become an essential tool for the study and analysis of complex systems such as financial markets. The structure of connections and dependencies intrinsic to commercial activities critically influences the economic landscape. As a complex system, financial markets can be modeled as networks, in which nodes represent individual investors and connections stand for interactions between them related to the purchase, sale or holding of financial assets. The convergence of individual decisions between different economic agents has repercussions on the behavior of the financial market observables, yielding essential collective behaviors, such as financial contagion and herd movements, speculative pricing, and crashes \cite{malkiel, Bachelier1900, BlackScholes1973, Mantegna2000}.

%The rich behavior and heterogeneity of a complex system as a society or a market can be graphically represented by networks or graphs, nodes in the network representing individuals, and the links between the nodes representing interactions between pairs of nodes. 
%In this sense, one can represent the interactions between financial agents on the market via a complex network framework. 

\begin{figure*}[ht]
	\centering
	\includegraphics[width=0.76\textwidth]{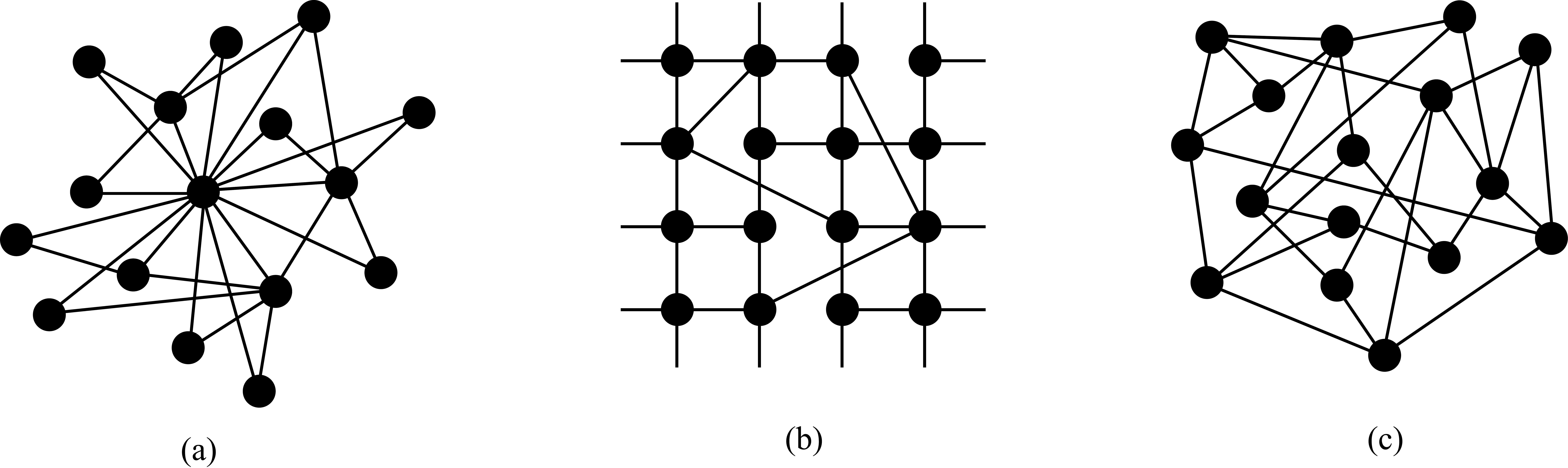}
	\caption{Illustration of distinct architectures of three complex networks with $N = 16$, and the average connectivity $\left< k \right> = 4$: from the hubs of (a) scale-free networks, (b) to the clusters and shortcuts of small-world networks and (c) the randomness of random networks.}
	\label{fig: cnillustration}
\end{figure*}

The dynamics of financial markets emerge from, among other factors, investors' rational and emotional activity, driven by the complex social dynamics and influences between agents in a network of economic investors. An example of such a phenomenon is herding, whereby individuals tend to follow the opinion or behavior of their neighbors. Buying, selling, or holding an asset is an agent's decision taken in a social environment with which he interacts. In this context, herding behavior has been suggested to play an essential role in finance, where often coherence in social and economic imitation manifests as informational cascades \cite{cont2000herd, Raafat2009, Hong2005, Zhao2011, Bikhchandani1992, Galam2008}.

In financial markets, behavioral finance has identified herding as a key to understanding the collective irrationality of investors \cite{Shiller2015}. An important class of economic agents is called \textit{noise traders}, who typically follow their neighbors' trends and tend to overreact to the arrival of news when buying or selling. Another essential group of agents seems to follow the trends of the global minority as an investment strategy. They tend to buy when noise traders drive prices down and sell when they move prices up. We shall refer to these agents as \textit{fundamentalists}, but they are also known as contrarians, sophisticated traders, or $\alpha-$investors \cite{Day1990, Voit, bornholdt2001expectation, kaizoji2002dynamics, takaishi2005simulations, lux1999scaling, Lux2000, DeLong1990}. For fundamentalist agents, the analysis of the fundamentals of an asset, based on financial data, guides their rational decision-making, and their action promotes price movements toward more realistic or \textit{fundamental} values. 

Over the years, several agent-based models capable of reproducing statistical features of real economic time series have been proposed as frameworks for understanding the dynamics of financial systems, with applications inspired by Ising systems and sociophysics models, such as the majority-vote dynamics \cite{Voit, bornholdt2001expectation, kaizoji2002dynamics, takaishi2005simulations, lux1999scaling, Lux2000, DeLong1990, sznajd2002simple, KKH2002, vilela2019majority, granha2022opinion, zubillaga2019three}. The agent-based majority-vote model enables the investigation of the time evolution of opinion in society using a state variable and social anxiety noise. Similar to the Ising model, it also exhibits second-order phase transitions in several network topologies for a critical noise level \cite{Brunstein1999, Tome2002, melo2010phase, vilela2020three, zubillaga2019three, granha2022opinion, bernardo2022, campos2003small, pereira2005majority, lima2007majority, lima2008majority, vilela2009, vieira2016phase, vilela2018effect}.

The inherent complexity of collective human behavior is subject to interdisciplinary considerations. However, microscopic models of opinion formation attempt to simplify group dynamics to essential interaction mechanisms. The majority-vote consensus dynamics capture critical social phenomena, inspiring scientists to expand it further to model group behavior and financial market evolution. In this context, the \textit{global-vote model} frames opinion dynamics as the foundation of agent decisions in financial markets \cite{vilela2019majority, zubillaga2019three, granha2022opinion}. The three-state model supports four essential features: individuals' strategies regarding market decisions, influence networks connecting agents in a market, a socioeconomic anxiety level, and agent financial action space. The latter is represented by a three-state stochastic variable standing for their opinion on a market decision, e.g., buying, selling, or holding an asset.

 In this work, we investigate economic opinion dynamics in financial markets by examining emergent collective economic phenomena in the global-vote market model on three distinct complex network topologies: scale-free, small-world, and random networks, and then comparing the outcomes \cite{ba-smcn, watts1998collective, erdos1960evolution}. We reveal insights into network topology's potential impact on market behavior by investigating how opinions propagate and evolve within these diverse network structures. Through our analysis, we seek to uncover valuable information about the interplay between market dynamics and network architecture, shedding light on the factors that may influence the formation and spread of financial opinions across different network types.
   
This work is organized as follows. Sec. \ref{findynamics} presents the main characteristics of the three-state global-vote model modeling for financial markets. Sec. \ref{numresults} exhibits our Monte Carlo numerical results on scale-free networks, small-world networks, and random graphs. In Sec. \ref{concfinrem}, we present our concluding remarks, briefly synthesizing the results of this work.

%=====================================================%
%=====================================================%

\section{Financial dynamics}
\label{findynamics}
We map the agent's financial decision at a given time $t$ by a stochastic variable, which may assume one of three states $s \in \{1, 2, 3\}$, which may represent buying, holding, or selling an asset. Financial market dynamics is driven by a heterogeneous composition of agents, randomly distributed in a network of social connections: a fraction $1 - f$ of noise traders and the remaining fraction $f$ of fundamentalist agents, also called noise contrarian traders. The former acts based on their nearest neighbors' decisions, whereas the latter on the behavior of the market as a whole. The total number of agents defines network size and equals $N$.

In the context of a society, two connected nodes may represent the fact that a pair of individuals know each other and may talk, trade, and exert influence on one another, for example. Different socioeconomic relations and organizational networks have a unique pattern of connectivity and structure. Figure \ref{fig: cnillustration} describes the visualization of three different complex networks with $N = 16$ nodes and average connectivity of $\left< k \right> = 4$ and highlights their fundamental properties. We illustrate (a) a scale-free network and its hub-like structure, (b) a small-world network characterized by the presence of clusters and shortcuts and (c) a random network with disordered connectivity. From network representation, we can infer fundamental principles that underlie complex systems and their collective critical behavior \cite{ba-smcn, watts1998collective, erdos1960evolution}.

Financial markets often reflect the socioeconomic stability of nations. Critical worldwide events may lead to economic anxiety and uncertainty, impacting stock market volatility. To model the level of economic anxiety present in a financial market, we introduce the socioeconomic noise parameter $q$. We assume that $q$ impacts both noise contrarians and noise traders' decisions and $q$ represents the probability of an agent not following its standard strategy when negotiating in the financial market \cite{vilela2019majority, granha2022opinion, zubillaga2019three}.

A noise trader agent updates its financial option according to the probabilistic prescription in Eq. (\ref{eq: noise trader prob}), following the three-state dynamics \cite{zubillaga2019three, melo2010phase, vilela2020three, bernardo2022, Brunstein1999, Tome2002}. A noise trader tends to follow the local majority, i.e., it agrees with the state of the majority of its nearest neighbors with probability $1 - q$ or dissents from it with probability $q$. Let $i$ represent a noise trader agent and $k_{i, s}$ represent the number of near-neighbors of $i$ occupying a given state $s \in \{1, 2, 3\}$. Below, we summarize the stochastic update rules for the state of a noise trader. The probability for an agent to adopt state $s = 1$ is given by

\begin{equation}
	\begin{array}{l}
		P\left(1|k_{i,1} > k_{i,2}; k_{i,3}\right) = 1 - q, \\
		P\left(1|k_{i,1} = k_{i,2} > k_{i,3}\right) = (1 - q)/2, \\
		P\left(1|k_{i,1} < k_{i,2} = k_{i,3}\right) = q, \\
		P\left(1|k_{i,1} ; k_{i,2} < k_{i,3}\right) = q/2, \\
		P\left(1|k_{i,1} = k_{i,2} = k_{i,3}\right) = 1/3.
	\end{array}
	\label{eq: noise trader prob}
\end{equation}

\noindent We remark that the probabilities for the remaining two states (2 and 3) follow from the symmetry operations of the $C_{3 \nu}$ group. Furthermore, the probabilities must be normalized, i.e., $P(1|\{ k_i \}) + P(2| \{ k_i \}) + P(3| \{ k_i \}) = 1$ for any global state configuration $\{ k_i \} \equiv \left\{ k_{i,1} , k_{i,2}, k_{i,3} \right \}$.

By contrast, a noise contrarian trader updates its financial option according to the probabilistic description in Eq. (\ref{eq: noise contrarian prob}). Fundamentalist agents tends to follow the global minority with probability $1 - q$ or dissents from it with probability $q$. Let $N_s$ represent the total number of agents in the network within a given state $s \in \{1, 2, 3\}$, where $N = N_1 + N_2 + N_3$. Below, we summarize the update rules for the state of a fundamentalist agent. The probability of adopting state $s = 1$ is

\begin{equation}
	\begin{array}{l}
		P\left(1|N_1 < N_2; N_3\right) = 1 - q, \\
		P\left(1|N_1 = N_2 < N_3\right) = (1 - q)/2, \\
		P\left(1|N_1 > N_2 = N_3\right) = q, \\
		P\left(1|N_1 ; N_2 > N_3\right) = q/2, \\
		P\left(1|N_1 = N_2 = N_3\right) = 1/3. 
	\end{array}
	\label{eq: noise contrarian prob}
\end{equation}

\noindent The probabilities for the remaining two states (2 and 3) follow from the symmetry operations of the $C_{3 \nu}$ group. The normalization condition, i.e. $P(1|\{ N \}) + P(2| \{ N \}) + P(3| \{ N \}) = 1$ also holds for any global state configuration $\{ N \} \equiv \left\{ N_1 , N_2, N_3 \right \}$.

Moreover, we remark that economic stability fundamentally impacts market volatility. Thus, the order parameter $M$ is defined in analogy to the three-state Potts model, and it measures the average market opinion, revealing the economic order

\begin{equation}
	M = \sqrt{M_1^2 + M_2^2 + M_3^2}.
\end{equation}
The order parameter $M$ will be referred to as the opinionization in analogy to the magnetization of physical spins. The opinionization measures the uniformity of opinion in the market. If all the agents share the same opinion, then $M = 1$. If the opinions are split evenly between the three states, then $M = 0$. We express $M$ as the magnitude of a vector with components

\begin{equation}
	M_s = \sqrt{\frac{3}{2}} \left ( \frac{N_s}{N} - \frac{1}{3} \right),
\end{equation}
with $s \in \{1, 2, 3\}$.

In the market context, we shall interpret the order parameter of the system $M$ as proportional to the price of a given asset \cite{bornholdt2001expectation, kaizoji2002dynamics, takaishi2005simulations}. We relate the time variations of the instantaneous opinionization $M(t)$ to a financial asset's logarithmic returns $r(t)$. Our model assumes that the investors' demands drive prices to update instantaneously. We define the logarithmic return at time $t$ as follows:

\begin{equation}
	r(t) = \textrm{log}\left[ M(t) \right] - \textrm{log}\left[ M(t-1) \right].
\end{equation}
The log-return measures the relative price changes of a financial asset between two instants of time. As such, it is a measure of the efficiency or performance of an investment. The volatility $v$ of a financial asset estimates the risk of investment in such an asset. A usual measure of volatility, locally in time, is the absolute value of the returns, since it quantifies the amplitudes of price variations as measures of fluctuations in the time series $v(t)\equiv|r(t)|$.

%=====================================================%
%=====================================================%

\section{Numerical results}
\label{numresults}
%=====================================================%
%=====================================================%

We perform Monte Carlo simulations on distinct complex network topologies with $N = 10^4$ nodes. We extensively investigate several network parameters: the growth parameter $z$ for scale-free networks, the rewiring parameter $p$ for small-world networks, and the average connectivity $\left<k\right>$ for random graphs. We build the network of financial interactions by considering a fraction $f$ of fundamentalist agents and the remaining fraction $1 - f$ as noise traders. From previous investigations \cite{vilela2019majority, zubillaga2019three, granha2022opinion}, we expect essential market features to emerge in the noise region near criticality $q \simeq q_c$, when contrarians are absent $f = 0$ for all networks investigated \cite{melo2010phase, bernardo2022, vilela2020three}.

At each instant or Monte Carlo step (MCS), we perform $N$ attempts to update the state of randomly selected individuals. Once selected, a financial agent updates his opinion accordingly with the probabilities given by Eqs. (\ref{eq: noise trader prob}) or (\ref{eq: noise contrarian prob}) for a noise trader or a fundamentalist agent, respectively. We randomize the initial state of the system, assigning to each agent any of the three available states with equal probability. We allow the dynamics to run during $10^4$ MCS to discard the transient regime and perform our analysis in the subsequent $10^4$ MCS. For every set of parameters $(q,f)$, we perform 100 Monte Carlo simulations for each network topology considered, averaging over network disorder. Therefore, a total of $10^6$ MCS was recorded for each pair of parameters $(q,f)$ and network topology from all the runs. In this way, the statistics gathered many realizations of the disorder caused by the random allocation of contrarians and the disorder from the network construction models.

In our investigation, we place $N$ agents on the nodes of three distinct network topologies: scale-free networks, small-world networks, and random graphs. Several network properties differentiate between each complex network topology \cite{ba-smcn}. The following sections detail the mechanisms for constructing the complex networks implemented in this work and the main results obtained for each topology considered. We aim to understand the socioeconomic dynamics under the effects of different network structures and compare our results with the observations made in behavioral economics and finance.

\subsection{Financial Markets on Scale-free Networks}

Upon analyzing the topology of social networks, airline networks, or the World Wide Web, we observe the presence of \textit{hubs} -- highly connected nodes -- a feature displayed by several other real-world networks. Such networks are frequently referred to as scale-free networks. The Barab{\'a}si-Albert model is a well-known method for building scale-free networks via two fundamental mechanisms: growth, where we consider that nodes are iteratively added, and preferential attachment, which describes the ``rich--get--richer'' effect, in which nodes with a high degree of connectivity have a higher chance of obtaining new connections \cite{barabasi2000scale, newman, ba-smcn}.

To generate scale-free networks, we use the Barab{\'a}si-Albert model and start with a fully connected core of $z$ nodes, where $z$ is the growth parameter. According to the preferential attachment algorithm, a new node adds $z$ new links to the existing network at each step of network growth. In this way, such networks present an average degree of $\left< k \right> = 2z$. We remark that the degree distribution of Barab{\'a}si-Albert scale-free networks display a power-law decay with exponent $\lambda \sim 3$ \cite{ba-smcn}.

 We focus the study of the financial dynamics near criticality $q \simeq q_c$ when there is an absence of contrarians since we expect opinionization fluctuations to be more significant in that region of the phase diagram. We collect the respective critical noise values, $q_c(z)$, from previous studies of the three-state majority-vote model on Barab\'asi-Albert networks \cite{vilela2020three}.

\begin{figure*}[ht]
	\centering
	\includegraphics[width=1\textwidth]{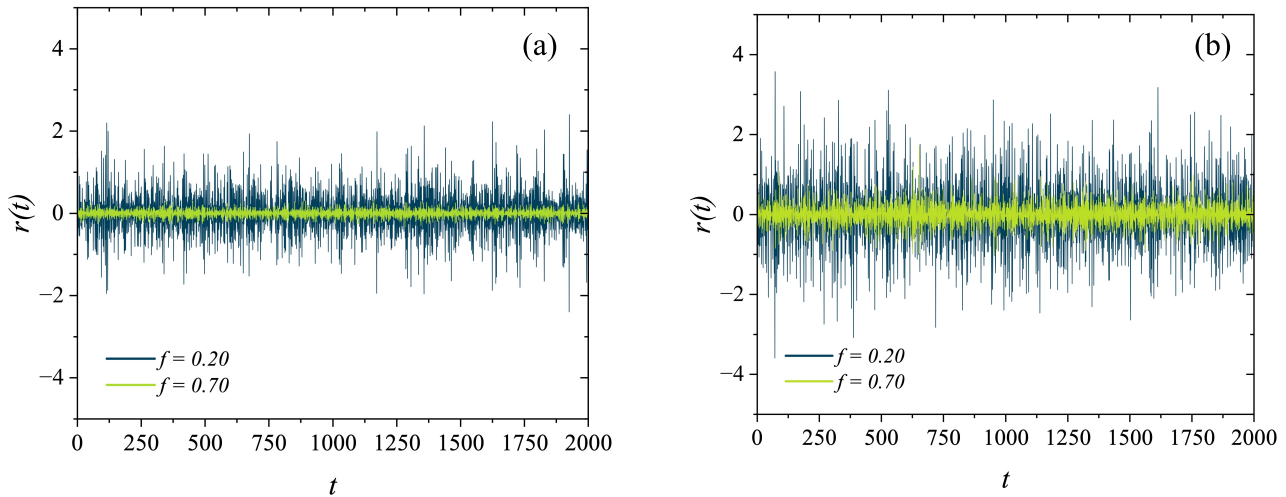}
	\caption{Time series of the logarithmic returns for (a) $z = 6$ and $q_c = 0.4550$, (b) $z = 50$ and $q = 0.5918$.}
	\label{fig: Returns-BA}
\end{figure*}

In Fig. \ref{fig: Returns-BA}, we show the influence of the growth parameter $z$ and the concentration of fundamentalists $f$ on the market behavior at the critical point. Fig. \ref{fig: Returns-BA}(a) illustrates two distinct market phases for $q_c = 0.4550$ and $z = 6$: a \textit{strong market phase} or \textit{turbulent phase} for $f = 0.20$, where the system presents several events of considerable volatility, as depicted by the large spikes in the plot; and a \textit{weak market phase} or \textit{laminar phase} for $f = 0.70$, where returns fluctuations are attenuated. This finding agrees with previous investigations in which increasing the number of contrarian agents drives market stability \cite{vilela2019majority, zubillaga2019three, granha2022opinion}. Furthermore, Fig. \ref{fig: Returns-BA}(b) shows that for $q_c = 0.5918$ and $z = 50$, the frequency of high volatility events for both values of $f$ increases with higher $z$, appearing in an uncorrelated form, indicating a deviation from the expected behavior of real-world financial markets.

Fig. \ref{fig: Returns-BA}(a) also demonstrates that periods of considerable return fluctuations are compressed for lower values of $f$, indicating the real-world market feature known as volatility clustering \cite{Lux2000, KKH2002, cont2007volatility}. This financial phenomenon can be comprehended by Mandelbrot's observation that ``large changes tend to be followed by large changes -- of either sign -- and small changes tend to be followed by small changes'' \cite{mandelbrot1963variation}. To quantify the effects of volatility clustering, we define the autocorrelation function of the absolute returns as follows:

\begin{equation}
	A(\tau) = \frac{\sum_{t = \tau +1}^{T} \left[\left|r(t)\right| - |\overset{-}{r}|\right]  \left[\left|r(t - \tau)\right| - |\overset{-}{r}|\right]}{\sum_{t = 1}^{T} \left[\left|r(t)\right| - |\overset{-}{r}|\right]^2},
	\label{Eq: ACR}
\end{equation}
where $1 \leq \tau \leq 10^4 \ \textrm{MCS}$ is the time-step difference between observations, $T = 10^4 \  \textrm{MCS}$ is the time of simulation for each network sample, $r(t)$ is the return at a time $t$ and $\overset{-}{r}$ the average return value.

\begin{figure*}[ht]
	\centering
	\includegraphics[width=1.0\textwidth]{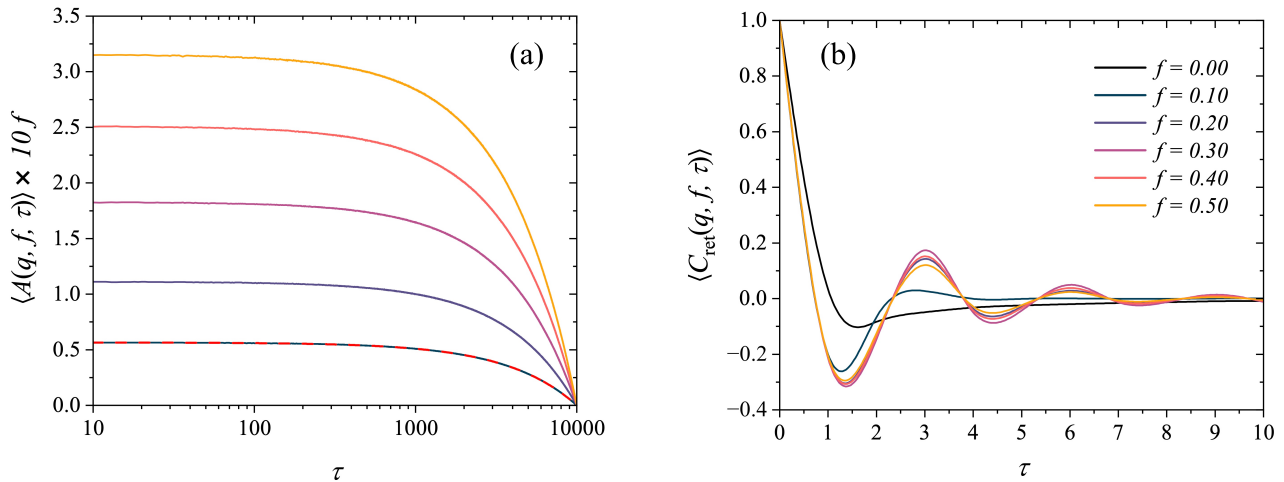}
	\caption{Averaged autocorrelation functions of the (a) volatility $\left| r(t) \right|$ and (b) logarithmic returns $r(t)$ for $z = 6$, $q_c = 0.4550$ and several values of $f$. The dashed red line corresponds to an exponential data fit.
	}
	\label{fig: ACR-z6-BA}
\end{figure*}

The function defined by Eq. (\ref{Eq: ACR}) measures non-linear correlations in a given time series, namely the autocorrelation function of the absolute value of log-returns, as a function of the time delay between observations. Many real-world studies demonstrate a strong positive correlation in the volatility $\left| r(t) \right|$ over extended periods such as days, weeks or months, consistent with the presence of volatility clustering in the data \cite{granha2022opinion}. Figure \ref{fig: ACR-z6-BA}(a) displays the average autocorrelation of the absolute value of log-returns for several values of $f$ and $100$ network samples. Results show the expressive correlation between high-volatility events and a slow exponential decay in time spanning over several orders of magnitude. We also exhibit an exponential fit of the data (dashed red line) $\left < A(q, f, \tau) \right > \sim exp(-\tau/\tau_0)$ for $f = 0.10$, in which we obtain $\tau_0 \approx 7 \times 10^7$ MCS \cite{Mantegna2000, kaizoji2002dynamics, Voit, gene1999}. Other values of $f$ yield similar results.

In Fig. \ref{fig: ACR-z6-BA}(b), we present the autocorrelation function $C_{ret}$ for the returns time series generated by the model for several values of the fundamentalist fraction $f$, averaged over $100$ network realizations. We connect data points using B-spline curves that smoothly interpolate between the control points, resulting in a smooth and continuous curve. As expected from real-world markets, returns are uncorrelated in time as $C_{ret} \to 0$ on a short-time scale of $\tau \sim O(1)$, reflecting that our model lacks long-term memory, agreeing with the efficient market hypothesis \cite{peiris2023revisiting, timmermann2004efficient, malkiel1989efficient}. For large fractions of contrarians, $C_{ret}$ displays diminishing anti-persistent oscillations in the short term. This oscillation stems from the cyclical dynamics of the system, primarily driven by a substantial portion of noise contrarians consistently striving to inhabit the instantaneous global minority state \cite{zubillaga2019three}.

\begin{figure}[h]
	\centering
	\includegraphics[width=0.48\textwidth]{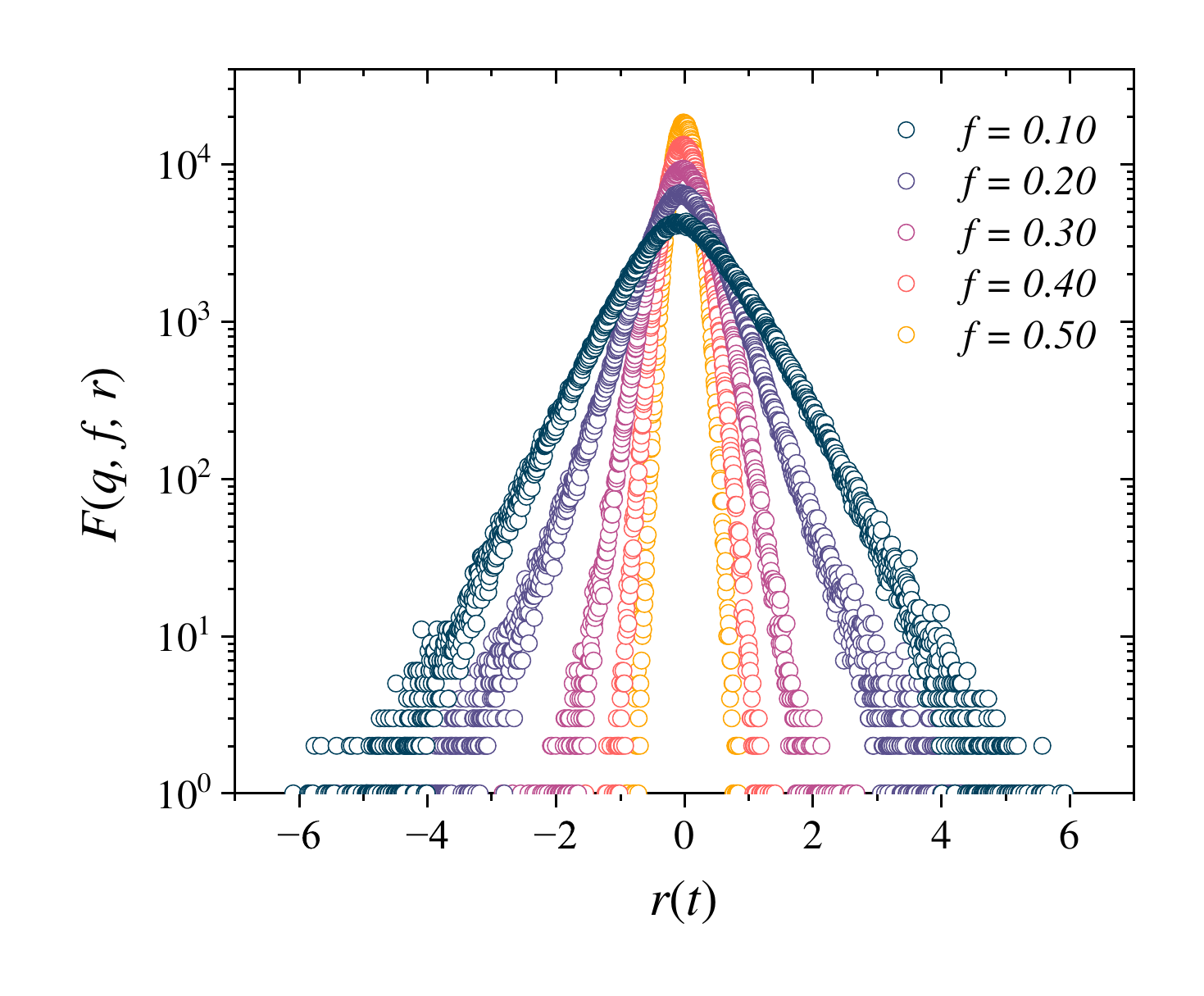}
	\caption{Distribution of logarithmic returns in $10^6$ MCS for $z = 6$ and $q = 0.4550$ and several values of the fraction of noise contrarians.}
	\label{fig: Histo-z6-BA}
\end{figure}

Figure \ref{fig: Histo-z6-BA} displays the histogram of the log-returns in $10^6$ MCS for $z = 6$ and several values of $f$. As previously discussed, we observe increased market stability for increasing fractions of contrarians in the network. Our results show that for lower values of the fraction $f$, the distributions display fat tails as a reflection of the high number of intense volatility events. As we increase the values of $f$, distributions' tails become less heavy as they gradually shift towards a Gaussian regime. To quantify the return distributions, we perform a statistical analysis of the data. We calculate the kurtosis $K(z, f)$ for the distributions and obtain $K(6, 0.10) = 4.48$ in the strong market phase and $K(6, 0.50) = 3.12$ in the weak market phase. Thus, increasing the fraction of contrarian agents $f$ shifts the system's behavior from a leptokurtic regime ($K > 3$) to a mesokurtic regime ($K \approx 3$).

\begin{figure}[h]
	\centering
	\includegraphics[width=0.48\textwidth]{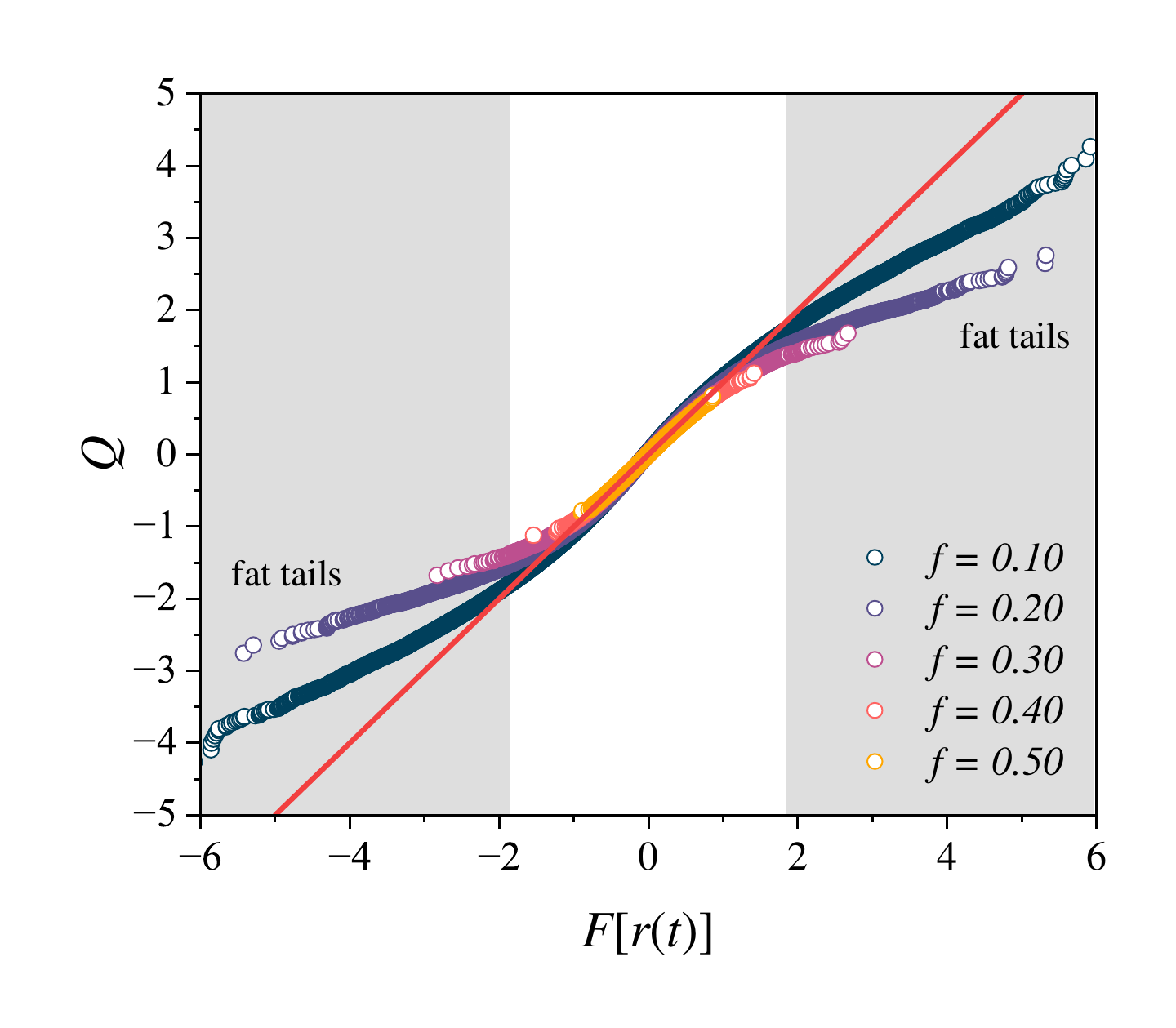}
	\caption{Normal quantile–quantile plots of the distributions of logarithmic returns in $10^6$ MCS for $z = 6$ and $q = 0.4550$. The red line displays the theoretical expected results for a Gaussian distribution.}
	\label{fig: QQplot-z6-BA}
\end{figure}

\begin{figure}[h]
	\centering
	\includegraphics[width=0.48\textwidth]{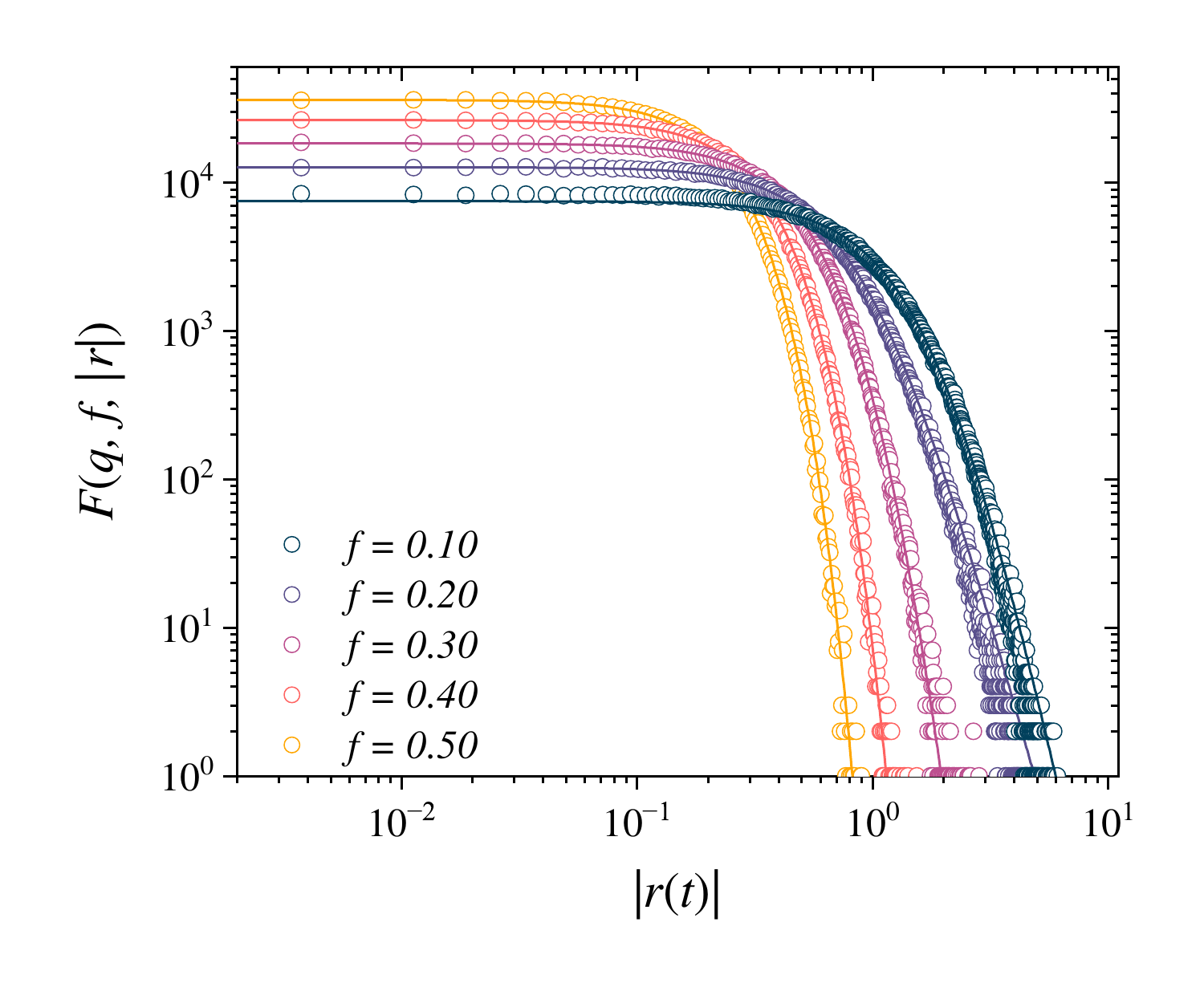}
	\caption{Distributions of the volatility for $z = 6$ and $q = 0.4550$ in $10^6$ MCS. The lines correspond to symmetric coupled exponential fits for the data.}
	\label{fig: student-fits-z6-BA}
\end{figure}

\begin{figure}[h]
	\centering
	\includegraphics[width=0.48\textwidth]{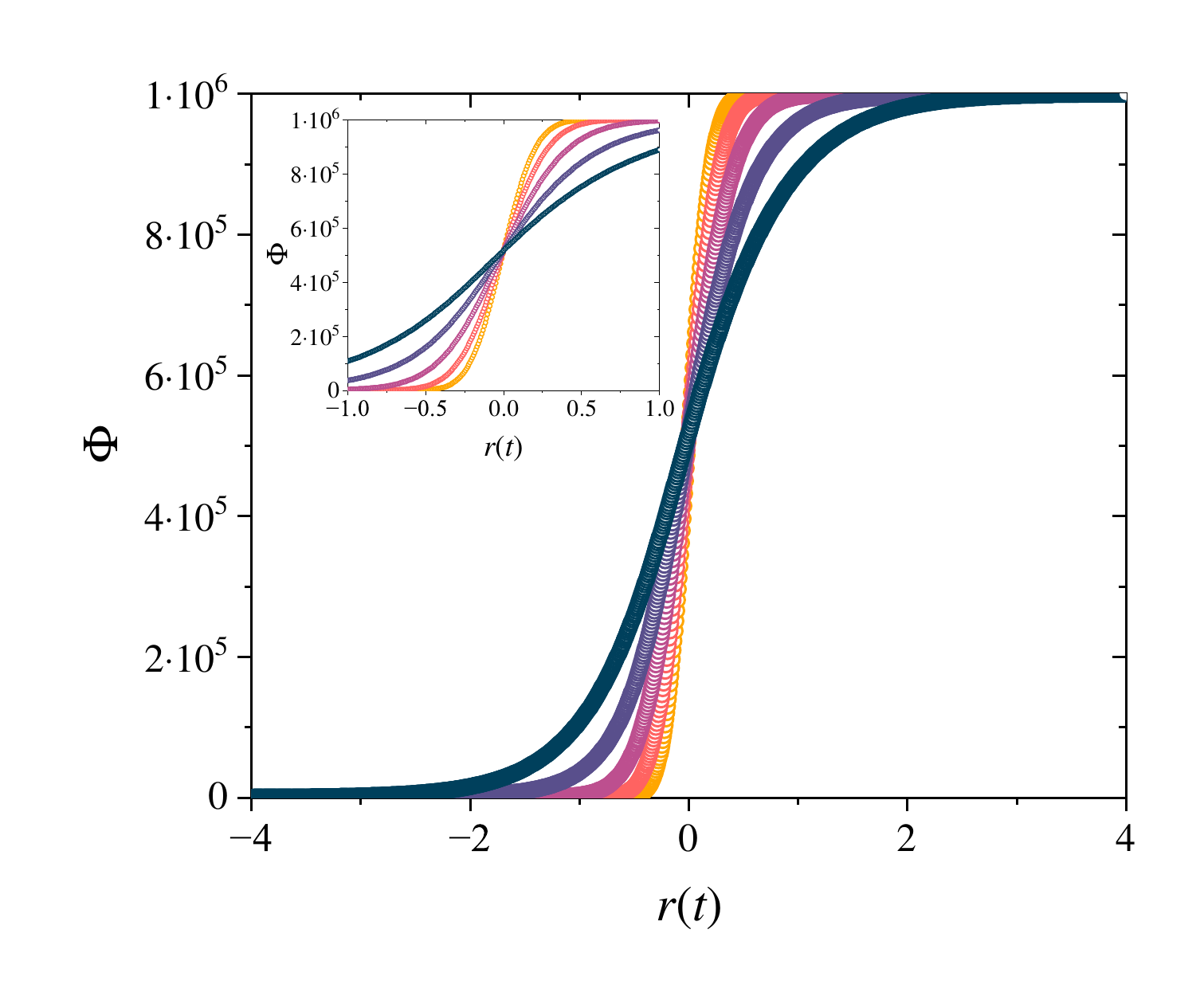}
	\caption{Cumulative distribution of logarithmic returns in $10^6$ MCS for $z = 6$ and $q = 0.4550$. The inset displays the behavior of $\Phi$ for $r(t)$ near zero.}
	\label{fig: cum-freq-z6-BA}
\end{figure}

To further qualify the distributions in Fig. \ref{fig: Histo-z6-BA}, we compute a comparative normal quantile-quantile (Q–Q) plot. In a Q-Q plot, the reference line, depicted in red, represents the expected results of a Gaussian distribution. If a particular distribution displays similar behavior, its data points should lie on that reference line. Thus Fig. \ref{fig: QQplot-z6-BA} displays the normal Q–Q plots for the distribution of log-returns and several values of the fraction of contrarians $f$. The fat-tailed behavior of the distributions for lower values of $f$, in particular $f \leq 0.30$, is observed via the nonlinearity in the Q–Q plots, thus indicating that the distributions are non-Gaussian and compatible with real-world financial data. Increasing the values of $f$ progressively shifts data points to the reference line, as distributions display a Gaussian behavior.

To quantify the transition of the log-return distributions from a heavy-tailed (leptokurtic) regime into a Gaussian (mesokurtic) regime, we consider the coupled exponential family of distributions $P_{\mu,\sigma,\kappa,\alpha}(r)$, where $\mu$ is the mean value, $\sigma$, $\kappa$ and $\alpha$ are the parameters of the function. We shall refer to the shape parameter $\kappa$ as the \textit{nonlinear statistical coupling} and to $\sigma$ as the scale parameter in an interpretation of non-extensive statistical mechanics of complex systems \cite{tsallis2003, tsallis2007, nelson2017, nelson2019, biondo2015}. Considering the mean value of the distributions as zero, in agreement with Fig. \ref{fig: cum-freq-z6-BA}, we fit the distributions via the symmetric coupled exponential family $P_{\mu,\sigma,\kappa,\alpha}(r) = P_{\sigma,\kappa,\alpha}(r)$, defined as follows.

\begin{equation}
    P_{\sigma,\kappa,\alpha}(r) \equiv \left[ Z(\sigma,\kappa,\alpha)\left(1+\kappa \Bigl|\frac{r}{\sigma} \Bigr|^\alpha \right)^{\frac{1+\kappa}{\alpha \kappa}}_{+} \right]^{-1},
    \label{eq: coupled exp fam}
\end{equation}
where $(x)_+ \equiv \max (0,x)$. Furthermore, if we set the parameter $\alpha = 2$, we obtain the coupled Gaussian distribution: for $\kappa = 0$, we recover the Gaussian distribution; and for $\kappa > 0$, we have the Student's t distribution, where the degree of freedom $\nu$ is related to the shape parameter as $\nu = 1/\kappa$. Thus, the shape parameter $\kappa$ yields a quantitative measure for the transition between distribution regimes \cite{tsallis2003, tsallis2007, biondo2015, nelson2017, nelson2019}.

\begin{table*}[ht]\centering
\ra{1.3}
\setlength{\tabcolsep}{0.25cm}
\caption{Correlation between the scale $\sigma$ and shape $\kappa$ as a function of the fraction of contrarian agents $f$, growth parameter $z$ and noise $q$.}
\begin{tabular}{@{}l l l l l l l@{}}

\\ \midrule
\ \ Fraction $f$ && 0.10 & 0.20 & 0.30 & 0.40 & 0.50 \ \  \\
\ \ Shape $\kappa$ && 0.16(3) & 0.205(2) & 0.070(2) & 0.024(2) & 0.017(1) \ \ \\
\ \ Scale $\sigma$ &&  0.74(1) \ & 0.4499(4) \ & 0.3204(3) \ & 0.2258(3) \ & 0.1651(1) \ \ \\

\bottomrule
\end{tabular}
\label{table: shape-scale}
\end{table*}

Figure \ref{fig: student-fits-z6-BA} displays the probability distribution of the volatilities for several values of the fraction of fundamentalists $f$ in $10^6$ MCS. The lines correspond to symmetric coupled exponential fits for the data, and the values obtained for the nonlinear coupling parameter $\kappa$ and scale $\sigma$ are portrayed in Table \ref{table: shape-scale}. The results provide a quantitative measurement of the gradual regime shift depending on the fraction of contrarians $f$: increasing $f$ pushes the nonlinear coupling parameter $\kappa$ towards zero, indicating the loss of fat tails as distributions become Gaussian; simultaneously, the scale $\sigma$ is progressively reduced, as the higher presence of fundamentalists tends to stabilize market dynamics. Thus, the coupled exponential fits performed provide a numerical measure of the transition from a heavy-tailed (leptokurtic) regime to a Gaussian (mesokurtic) regime depending on the fraction of contrarians $f$.

\begin{figure*}[ht]
	\centering
	\includegraphics[width=0.8\textwidth]{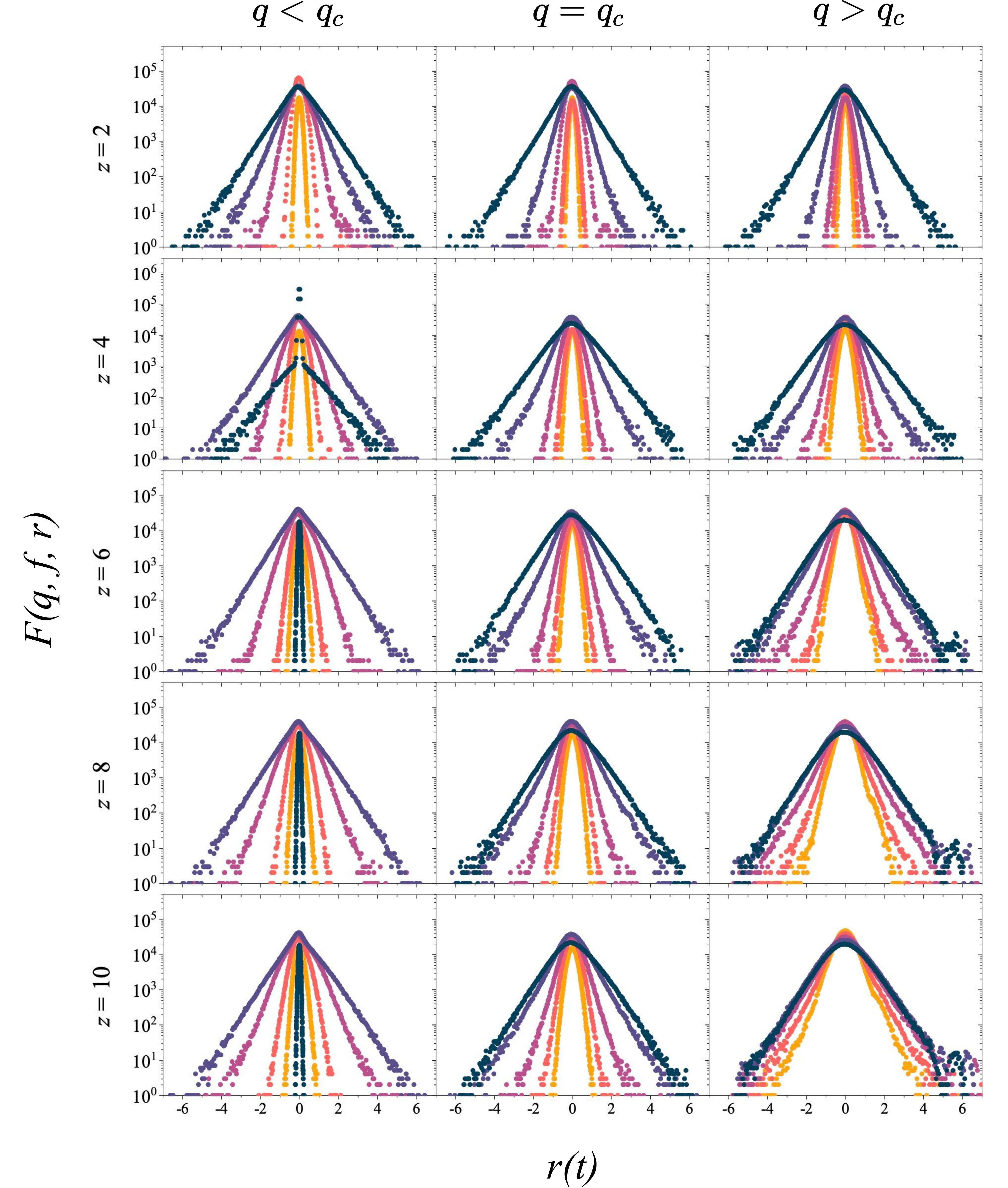}
	\caption{Distributions of logarithmic returns in $10^6$ MCS for different combinations of parameter values ($z, q$) in the vicinity of $q_c (z)$ with  $f = 0.10, \ 0.20, \ 0.30, \ 0.40,$ and $0.50$ in dark blue, purple, pink, orange, and yellow, respectively.}
	\label{fig: Histo-combined-BA}
\end{figure*}

Figure \ref{fig: cum-freq-z6-BA} displays the cumulative distribution of logarithmic returns $\Phi$ for $z = 6$ in $10^6$ MCS. As previously stated, we observe that the mean of the distributions remains zero for all investigated values of $f$.

We also extensively investigate the model's behavior for several micro-state configurations, depending on the growth parameter $z$, the fraction of contrarians $f$, and the noise parameter $q$. Figure \ref{fig: Histo-combined-BA} displays a multi-plot table of the histograms of logarithmic returns for several ($q, f, z$) triplets investigated near the criticality of the system $q \approx q_c(z)$. Columns correspond to values of $q$, below criticality (left), at criticality (center), and above criticality (right). The selected noise values above and below criticality are $q = q_c(z) \pm 0.1$.

As one moves vertically (from top to bottom) in the grid along a column, one observes the topological effect of increasing the growth parameter of the network, especially in the center column. Increasing $z$ reveals that the spreads and tails of the return distributions tend to increase slightly, suggesting broader return distributions. Furthermore, shifting towards higher values of $f$ leads to a progressive loss of tails in the return distributions. In contrast, this behavior is lost for noise parameter values that deviate from criticality, as observed in particular by the distributions for $f = 0.10$ below criticality, which becomes highly peaked around $r = 0$ for higher values of $z$, as well as the distributions for $z= 8$ and $10$ above criticality, where higher fractions of contrarians still display a heavy-tailed behavior. Hence, this result conforms with our choice for the growth parameter $z$ and noise parameter $q = q_c$ values.

\begin{figure*}[ht]
    \centering
    \includegraphics[width=1\textwidth]{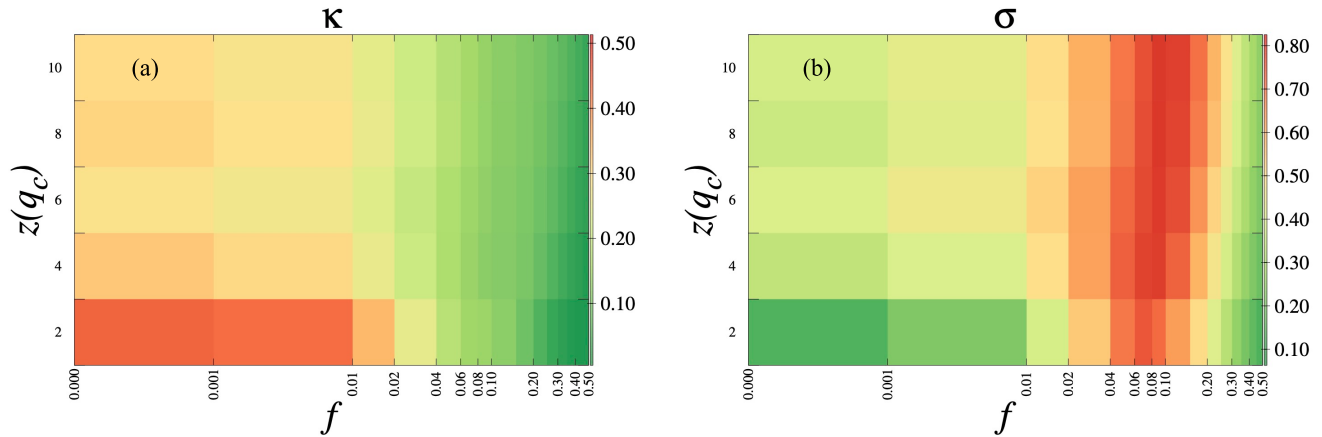}
    \caption{Heat map of (a) the nonlinear statistical coupling $\kappa$ and (b) the scale parameter $\sigma$ as a function of the critical noise parameter $q_c(z)$ at each corresponding value of $z$ and the percentage of contrarians $f$.}
    \label{fig:BAheatmaps}
\end{figure*}

Furthermore, we consider a quantitative approach to the distributions shown in Fig. \ref{fig: Histo-combined-BA}, focusing exclusively on the system's behavior at the critical point $q = q_c(z)$. We perform fits for the volatility distributions of the system for different growth parameter values at their corresponding critical points according to the symmetric coupled exponential family of distributions Eq. (\ref{eq: coupled exp fam}) with $\alpha = 2$. Thus, Fig. \ref{fig:BAheatmaps} displays a heat map of the fitting parameters obtained for the nonlinear coupling $\kappa$ and the scale parameter $\sigma$ for several values of $z$. The rows correspond to different average connectivity values, each at their own critical noise parameter. Along the rows, we explore the effect of the different values of the fraction of contrarians $f$. 

Figure \ref{fig:BAheatmaps}(a) displays the heat map for $\kappa$ where we observe the usual transition from a leptokurtic to a mesokurtic regime with increasing fractions of contrarians, evidence of the loss of heavy tails tending toward a Gaussian distribution for high values of $f$. For large enough values of $f$, this transition occurs in a universal-like way for Barab\'asi-Albert networks. Nevertheless, we observe the effect of the topology on the tails of the distributions for small values of $f$. Indeed, the distributions for $z=2$ display considerably heavier tails when compared with higher values of $z$.  

Furthermore, Fig. \ref{fig:BAheatmaps}(b) displays the heat maps for the scale parameter $\sigma$. We note that the scale parameter is small for $f\approx 0$. It rapidly grows with $f$, reaching a peak around the same value of $f$, decaying for larger fractions of contrarians as a reflection of the aforementioned transition between leptokurtic and mesokurtic regimes. Furthermore, we observe the effects of the topology on the volatility distributions as the peaks shift slightly to higher values of $f$ as $z$ increases.

We now proceed to investigate the influence of small-world networks and Erd{\"o}s-R\'enyi random graphs on the behavior of the system's dynamics. For these topologies, the lack of correlations in the time series of log-returns is still present, consistent with the efficient market hypothesis. Anti-persistence emerges for large values of contrarians. Long-term correlation is still a feature of the volatilities as a reflection of the volatility clustering effect. Therefore, in the following subsections, we shall explore the effects of the topology on the histograms of log-returns.

\subsection{Small-world Effects on Financial Dynamics}

The term small-world effect alludes to the fact that most pairs of nodes in many real-world networks connect their elements by paths of short lengths, even though the sizes of complex networks are typically very large \cite{watts1998collective}. Furthermore, small-world networks encompass two essential features of real-world networks: short average path lengths and many network cliques. 

Inspired by the Watts-Strogatz model, we build our networks by rewiring the links that connect the $N = L \times L$ nodes of bidimensional square lattices with probability $p$, while forbidding rewiring to the original nearest neighbors and double connections \cite{watts1998collective, campos2003small, bernardo2022, zubillaga2019three}. We refer to $p$ as the rewiring parameter, which also relates to the degree of randomness of the rewired network. When $p = 0$, we recover the standard square lattice, whereas we obtain a random network for $p = 1$. We remark that the rewiring process does not affect the original average connectivity of the square network. Thus, for small-world networks built via rewiring square lattices, $\left< k \right > = 4$ for all values of $p$ \cite{watts1998collective, zubillaga2019three}.

We examine the return distributions and shall focus our investigation on the region around the critical points $q_c(p)$ for each corresponding value of the rewiring parameter $p$ to investigate distinct system configurations near criticality. In this case, we also reference the critical social anxiety level $q_c(p)$ in market dynamics, with values obtained from previous investigations of the three-state majority-vote model on small-world networks \cite{zubillaga2019three}.

\begin{figure*}[ht]
    \centering
    \includegraphics[width=0.8\textwidth]{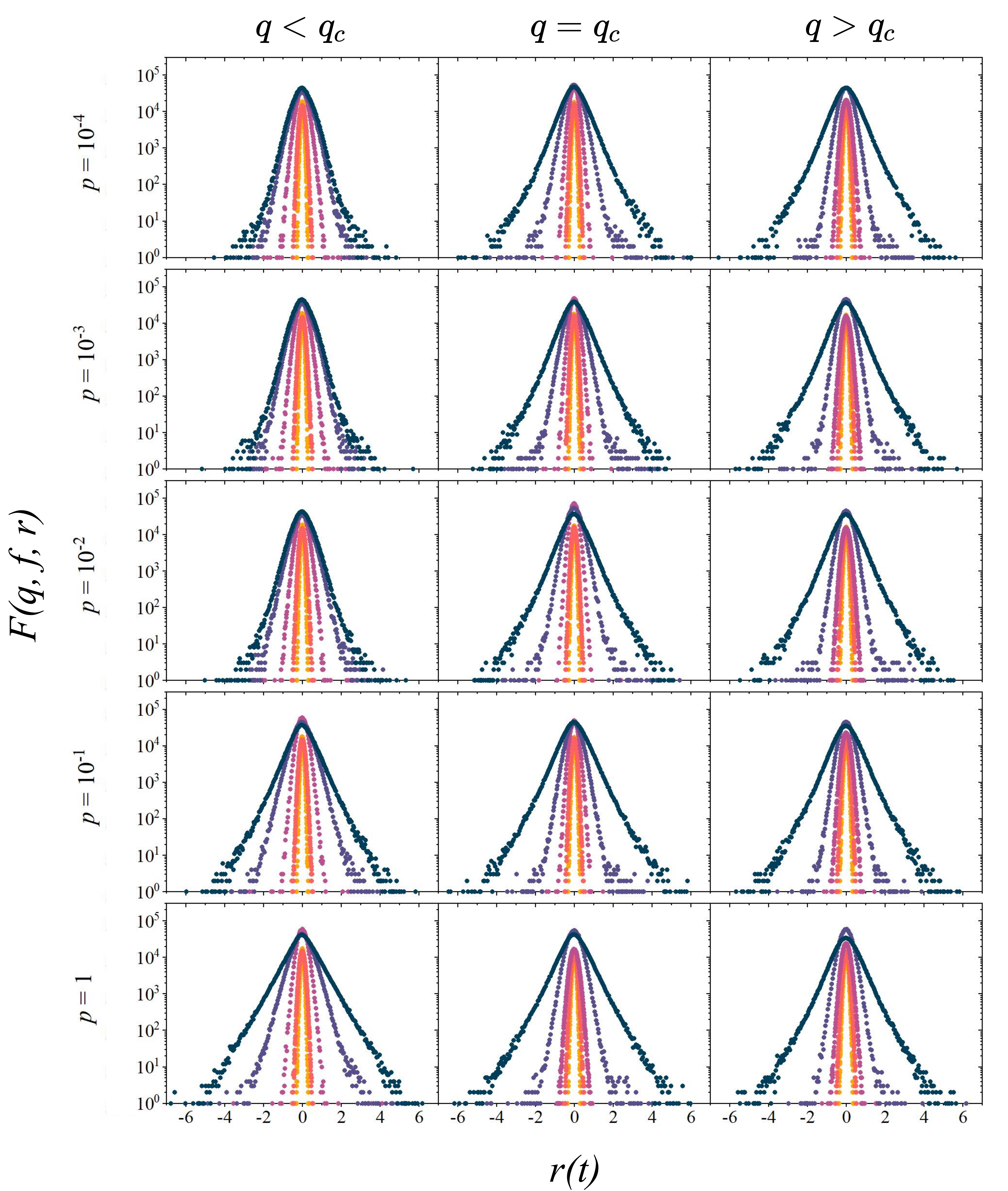}
    \caption{Distributions of logarithmic returns for various noise parameters $q$ and rewiring probability $p$ in the vicinity of $q_c(p)$ for several values of the fraction of contrarians: $f = 0.10, \ 0.20, \ 0.30, \ 0.40,$ and $0.50$ (dark blue, purple, pink, orange, and yellow, respectively).}
   \label{fig:SWreturns}
\end{figure*}

In Fig. \ref{fig:SWreturns}, we present a multi-plot table with the logarithmic return distributions for several $(q,f,p)$ triplets. In analogy to the previously performed investigation, the rows correspond to distinct rewiring probabilities $p$, whereas the columns correspond to noise values below criticality, at criticality and above criticality for each value of $p$ considered. Hence, each cell in the grid depicts a $(p,q)$ pair and shows several distributions corresponding to different concentrations of contrarians. The values of noise above and below criticality are taken to be $q \equiv q_c \pm 0.07$.

We observe that for macroscopically relevant fractions of contrarians, an increase in $f$ leads to a progressive loss of tails in the return distributions, suggesting a similar transition between a leptokurtic regime and a mesokurtic regime. This feature is displayed in all the cells in the grid, i.e., for a broad spectrum of small-world networks at criticality or in its vicinity. Thus, the model shows evident robustness for these networks over different topologies and social temperatures $q$ (above and below the critical point). Furthermore, moving vertically (from top to bottom) in the grid along a column displays the topological effect of increasing the rewiring probability of the system. In particular, for $q < q_c$, increasing $p$ reveals that the spreads and tails of the return distributions also tend to increase, especially for low values of $f$. In contrast, the behavior along the $q = q_c$ and $q > q_c$ columns suggests that not much variation (if any) is present as one varies $p$, suggesting a universal-like behavior. 

\begin{figure*}[ht]
    \centering
    \includegraphics[width=1.0\textwidth]{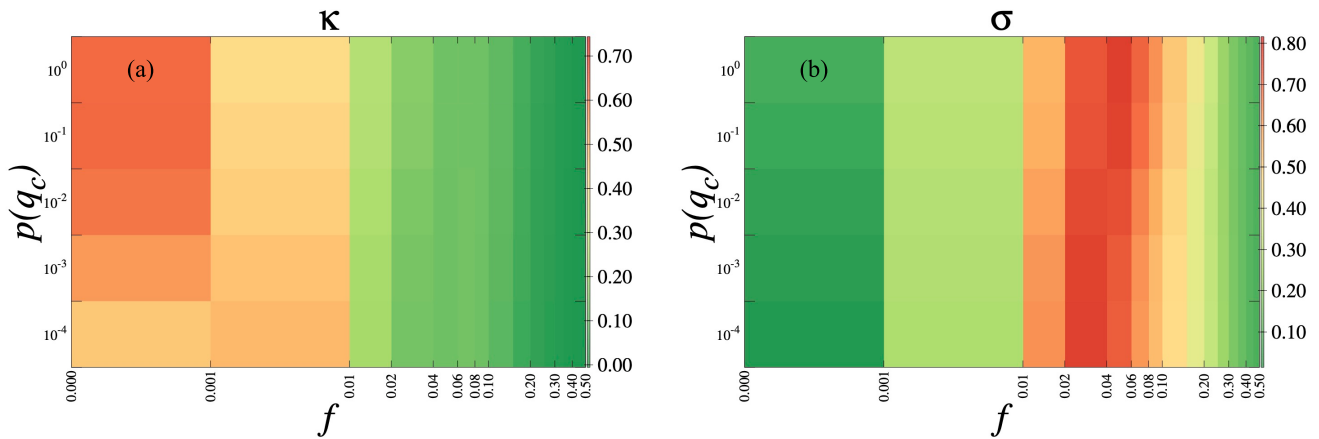}
    \caption{Heat map of (a) the nonlinear statistical coupling $\kappa$ and (b) the scale parameter $\sigma$ as a function of the critical noise parameter $q_c(p)$ at each corresponding rewiring probability $p$ and the fraction of contrarians $f$.}
    \label{fig:SWheatmaps}
\end{figure*}

We follow the above discussion with a quantitative approach, focusing on the volatility distributions for different values of $p$ at their corresponding critical points $q=q_c(p)$. We perform fits for the data according to the symmetric coupled exponential family of distributions Eq. (\ref{eq: coupled exp fam}) with $\alpha=2$. In Fig. \ref{fig:SWheatmaps}, we display the fitting parameters corresponding to the nonlinear coupling (a) $\kappa$ and the scale parameter (b)n $\sigma$ by means of heat maps. The rows correspond to distinct values of $p$, and along the rows, we explore the effects of different values of the fraction of contrarians $f$.

Once more, Fig. \ref{fig:SWheatmaps}(a) displays the clear transition from a leptokurtic to a mesokurtic regime, i.e., the loss of tails due to the increasing fractions of contrarians. As previously remarked, for large enough values of $f$, this transition takes place universally for small-world networks built according to the link rewiring scheme. Fig. \ref{fig:SWheatmaps}(b) shows the heat map corresponding to the scale parameter $\sigma$. Following the discussion regarding random networks, the scale parameter is small for $f\approx 0$, rapidly increasing with $f$, reaching a peak and then decaying similarly for macroscopically large fractions of contrarians. This indicates that the tails are lost in this regime as the distributions approach a Gaussian behavior.

\subsection{Random Interacting Networks}
A random network connects pairs of nodes with chance $0 < w \leq 1$, having an expected number of links equal to $\left < L \right > = wN(N-1)/2$ out of the maximum number of links $L_{max} = N(N-1)/2$ as $w \to 1$. We adopt the Erd{\"o}s-R\'enyi method for enabling random graphs on computers. This method adds $wN (N - 1)/2$ links to the $N$ initially isolated nodes, forbidding double connections. Such networks display a Poisson degree distribution for large values of $N$, with an average degree of connectivity $\left < k \right > = 2\left < L \right >/N = w (N - 1)$ or $\left < k \right > \approx wN$ \cite{erdos1960evolution, bollobas1985random, newman}. We study market evolution under a socioeconomic network of randomly connected opinions and average connectivity $ \left < k \right >$ as a function of the anxiety level $q$ and the fraction $f$ of fundamentalist agents.

\begin{figure*}[ht]
    \centering
    \includegraphics[width=0.8\textwidth]{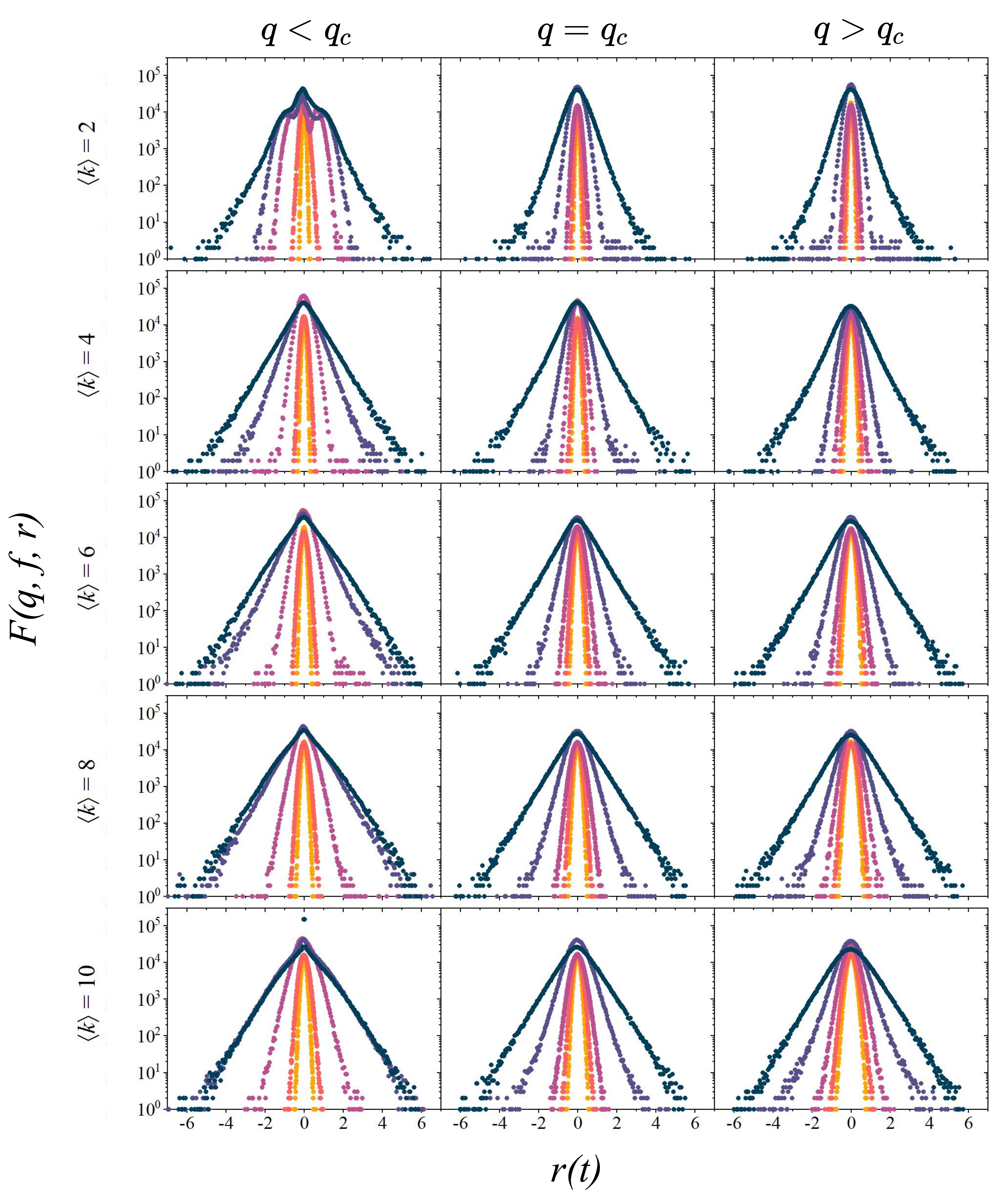} 
    \caption{Distributions of logarithmic returns for various noise parameters $q$ and average connectivity $\left< k \right>$ in the vicinity of $q_c(\left< k \right>)$ for various concentrations of contrarians: $f = 0.10, \ 0.20, \ 0.30, \ 0.40,$ and $0.50$ (dark blue, purple, pink, orange, and yellow, respectively).}
   \label{fig: RG histo returns}
\end{figure*}

In Fig. \ref{fig: RG histo returns}, we present a multi-plot table with a set of logarithmic return distributions for several $(q, f, \left< k \right>)$ triplets. In the plot, rows correspond to simulations done on random networks with different values of $\left< k \right>$, whereas the three columns represent values of the noise parameter $q$ below criticality, at criticality, and above criticality for each studied value of $\left< k \right>$. In this way, each cell in the grid relates to a pair of values $(\left< k \right>, q)$ and investigates several system micro-states regarding the fraction of contrarians $f$. The noise values above and below criticality are $q \equiv q_c \pm 0.1$, as we explore the model dynamics near criticality.

As in the other topologies explored thus far, it is clear from Fig. \ref{fig: RG histo returns} that an increase in $f$ leads to the aforementioned loss of tails in the return distributions, indicating the gradual shift between leptokurtic and mesokurtic regimes. We observe such behavior for most cells in the grid, suggesting that the model is robust over a wide range of topologies and noise parameter values. Nevertheless, we remark on some visible exceptions, such as the distributions for $\left< k \right> = 2$ below criticality, where lower fractions of contrarians display unusual behavior. Furthermore, moving vertically (from top to bottom) along a column sheds light on the topological effects of increasing $\left< k \right>$. In this way, increasing the growth parameter of the network correspondingly increases slightly the spreads and tails of the return distributions, as observed in previous investigations of the two-state global-vote model \cite{granha2022opinion}.

\begin{figure*}[ht]
    \centering
    \includegraphics[width=1.0\textwidth]{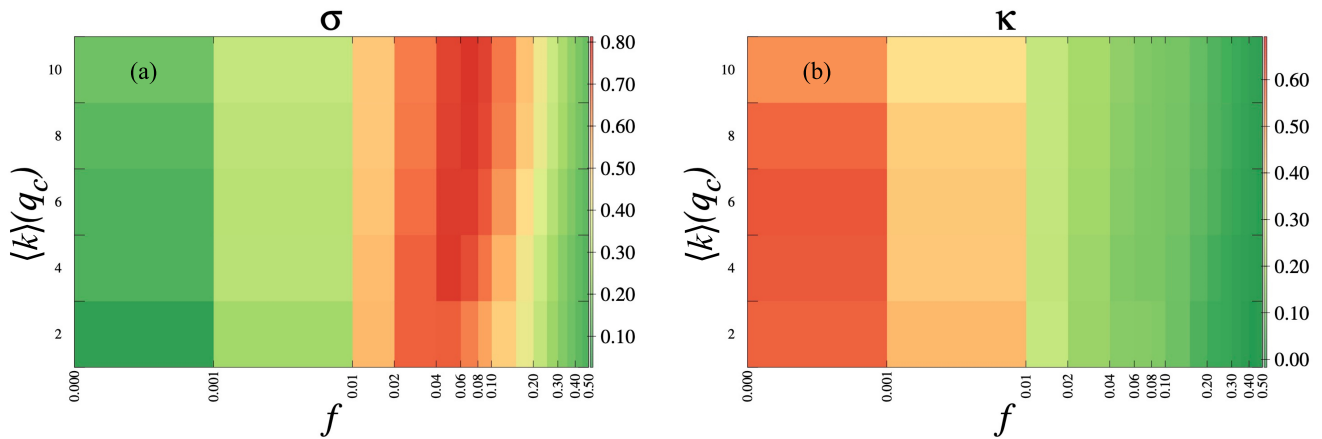}
        \caption{Heat map of (a) the nonlinear statistical coupling $\kappa$ and (b) the scale parameter $\sigma$ as a function of the critical noise parameter $q_c(\left< k \right>)$ at each corresponding average degree $\left< k \right>$ and the fraction of contrarians $f$.}
    \label{fig:ERheatmaps}
\end{figure*}

We deepen the previous discussion with a quantitative approach by focusing on the system's behavior at the critical point $q=q_c(\left< k \right>)$, as we did for scale-free and small-world topologies. We fit the volatility distributions of the system for different values of $\left< k \right>$ according to the symmetric coupled exponential family of distributions Eq. (\ref{eq: coupled exp fam}) with $\alpha=2$. Figure \ref{fig:ERheatmaps} displays the heat maps of the fitting parameters corresponding to the nonlinear coupling $\kappa$ and the scale parameter $\sigma$. The rows correspond to distinct values of the average connectivty $\left< k \right>$, and we explore the influence of different values of the fraction of contrarians $f$ along the rows.

Fig. \ref{fig:ERheatmaps}(a) depicts the gradual transition from a leptokurtic to a mesokurtic regime with increasing fractions of contrarians, evidence of the loss of heavy tails for all investigated values of $\left< k \right>$. Similarly, Fig. \ref{fig:ERheatmaps}(b) displays the scale parameter's behavior for different system configurations: $\sigma$ is small for small values of $f \approx 0$, rapidly growing with $f$; it peaks around the same value of $f$ before decaying for macroscopically large fractions of contrarians. The peak, however, shifts slightly to higher values of $f$ as $\left< k \right>$ increases, a manifestation of the effect of increasing the average degree. The scale parameter of the distribution will then tend to decay for larger values of $f$ as the tails weaken and the distributions approach a Gaussian behavior.

\begin{figure*}[ht]
    \centering
    \includegraphics[width=1.0\textwidth]{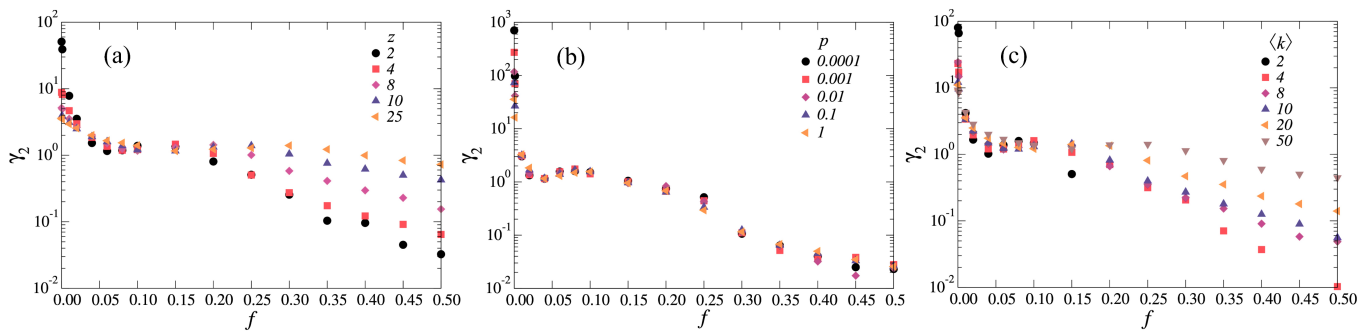} 
    \caption{Excess kurtosis of the return distributions for several values of the fraction of contrarians $f$ and different values of the network parameters at their corresponding critical noise values (a) $z$ for scale-free networks, (b) $p$ for small-world networks and (c) $\langle k \rangle$ for random networks. Recall that the excess kurtosis of a Gaussian distribution is $\gamma_2 = 0$.}
   \label{fig: kurtosisBASWER}
\end{figure*}

Additional quantification of the transition between fat-tailed and normal regimes is obtained via calculating the excess kurtosis of the return distributions, where for $\gamma_2 = 0$, we recall a Gaussian distribution. Thus, in Fig. \ref{fig: kurtosisBASWER}, we plot the excess kurtosis of the return distributions for several values of the fraction of contrarians and different values of network parameters, (a) the growth parameter $z$ for scale-free networks, (b) the rewiring probability $p$ for small-world networks and the average connectivity (c) $\langle k \rangle$ for random networks, at their corresponding real-world market (critical) noise values. We confirm that higher values of $f$ tend to soften the distribution tails for all investigated network topologies, driving the market into a Gaussian regime.

Furthermore, closer inspection of Fig. \ref{fig: kurtosisBASWER} reveals the effects of the network’s average connectivity on the distribution of logarithmic returns of the model. Recall that the average degree for scale-free networks is $\langle k \rangle = 2z$ whereas $\langle k \rangle = 4$ is constant and independent of the rewiring probability $p$. Thus, Fig.  \ref{fig: kurtosisBASWER}(a) and (c) show that the kurtosis’ decay tends to spread out for increasing values of $f$ as we vary the network’s average degree of connectivity. We observe that high (low) values of the average individual connectivity promote a slow (rapid) decay of $\gamma_2$. Therefore, the network’s average degree plays a crucial role in shaping the distribution of returns.

%=====================================================%
%=====================================================%
\section{CONCLUSION AND FINAL REMARKS}
\label{concfinrem}

This work investigates the stochastic dynamics of a three-state economic opinion formation model on complex networks, extending and generalizing previous investigations on the three-state global-vote model for financial markets \cite{vilela2019majority, zubillaga2019three, granha2022opinion}. As in the standard version, two different types of individuals are considered: noise traders who interact locally with their nearest neighbors and tend to agree with the state of the \textit{local majority} with probability $1-q$ and fundamentalists, who are subject to global interactions with the market as a whole and tend to follow the state of the \textit{global minority} with probability $1-q$. The parameter $q$ quantifies the socioeconomic anxiety level.

Financial agents are represented as nodes on complex networks, and the links between neighboring pairs of nodes represent opinion-driven financial interactions. We simulate the dynamics of the model on scale-free, small-world, and random networks and investigate how the distributions of returns are influenced by modifying specific network parameters, namely the growth parameter $z$ for scale-free networks, the rewiring parameter $p$ for small-world networks and the average connectivity $\left < k \right >$ for random networks. In this work, financial systems comprise three main features: a heterogeneous population of agents with distinct strategies, a complex network of financial interactions, and a level of economic uncertainty near some consensus-dissensus criticality.

By relating changes in the instantaneous financial order of this system to price fluctuations, the model can reproduce the main features of real-world financial markets \cite{bornholdt2001expectation, kaizoji2002dynamics, takaishi2005simulations}. Our results display such stylized facts of financial time series as fat-tailed distributions of returns, volatility clustering, and long-term memory of the volatility, consistent with the efficient market hypothesis and previous investigations \cite{vilela2019majority, zubillaga2019three, granha2022opinion}.

The logarithmic returns of the simulations fit a coupled exponential distribution, which is parameterized by the scale or generalized standard deviation and the shape or nonlinear statistical coupling \cite{tsallis2003, tsallis2007, nelson2017, nelson2019, biondo2015}. This family of distributions is typically used within the context of non-extensive statistical mechanics as a tool for the characterization of the complexity of a system. For macroscopically relevant fractions of contrarians, an increase in the contrarians decreases both the scale and the shape of the distributions. This macroscopic effect results from a loss of local order at a microscopic level and the emergence of global order, deviating the dynamics from heavy-tailed real-world to Gaussian distributions of returns.

We frame socioeconomic anxiety levels with the critical point where the opinionization fluctuations diverge in the thermodynamic limit, leading to return distributions with heavy tails and volatility clustering. Furthermore, the topology effects of the underlying network on the returns and volatilities are observed in the distinct ways that the tails decay for networks with different average degrees. The higher the average degree of the network, the slower the decay of the tails of the distributions with increasing fractions of contrarians, as evidenced by the behavior of the performed fits for scale-free and random networks. In contrast, since the average degree for the small-world networks is constant $\langle k\rangle=4$, the decay of the tails appears to approach a universal behavior independent of the rewiring probability $p$. 

We observe that the larger the fraction of contrarians, the less important the role of the topology is, and the more mean-field the financial system becomes. This observation is reasonable since contrarians do not interact locally with their neighbors. In fact, they interact globally with the state of the market as a whole. Our findings suggest that the behavior of the model is closely tied to broader factors such as the average connectivity of the analyzed network, the level of socioeconomic anxiety within it, and the proportion of fundamentalist agents, regardless of the detailed topology of the socioeconomic networks.

\begin{center}
\textbf{Acknowledgements}
\end{center}

We acknowledge financial support from POLI-UPE, and the funding agencies FACEPE (APQ-0565-1.05/14, APQ-0707-1.05/14), CAPES and CNPq (306068/2021-4). The Boston University work was supported by NSF Grants PHY-1505000, CMMI-1125290, and CHE-1213217, by DTRA Grant HDTRA1-14-1-0017, and by DOE Contract DE-AC07-05Id14517.

\begin{center} 
\textbf{ADDITIONAL INFORMATION}
\end{center}

{\textbf{Author Contribution statement}}

\noindent 
Bernardo J. Zubillaga, Mateus F. B. Granha, Andr\'e L. M. Vilela, Chao Wang: Conceptualization, Methodology, Software, Validation, Investigation, Formal analysis, Writing. Kenric P. Nelson: Investigation, Formal analysis, Writing. H. Eugene Stanley: Formal analysis, Writing.

{\textbf{Competing Financial Interests statement}}

\noindent
The authors declare no competing financial and non-financial interests.

\end{document}